\documentclass[preprint,aps]{revtex4}
\usepackage{graphicx}
\begin{document}
\title{Modeling one-dimensional island growth with mass-dependent detachment
rates}
\author{R. B. Stinchcombe${}^{1,}$\footnote{E-mail address:
r.stinchcombe1@physics.ox.ac.uk} and F. D. A. Aar\~ao
Reis${}^{2,}$\footnote{Email address: reis@if.uff.br}}
\affiliation{
${}^{1}$ Rudolf Peierls Centre for Theoretical Physics, Oxford University, 1
Keble Road, Oxford OX1 3NP\\
${}^{2}$ Instituto de F\'\i sica, Universidade Federal Fluminense, Avenida
Litor\^anea s/n, 24210-340 Niter\'oi RJ, Brazil
}
\date{\today}
\begin{abstract}
We study one-dimensional models of particle diffusion and attachment/detachment
from islands where the detachment rates $\gamma$ of particles at the cluster
edges increase with cluster mass $m$. They are expected to mimic the effects of
lattice mismatch with the substrate and/or long-range repulsive
interactions that work against the formation of long islands. Short-range
attraction is represented by an overall factor
$\epsilon\ll 1$ in the detachment rates (masses $m\geq 2$) relatively to
isolated particle hopping rates ($\epsilon\sim e^{-E/T}$, where $E$ is a binding
energy and $T$ is the temperature). We consider various mass-dependences of
$\gamma (m)$, from rapidly increasing forms such as $\gamma(m)\sim m$ to
slowly increasing ones, such as $\gamma (m)\sim {\left(\frac{m}{m+1}\right)}^b$,
with constant $b$. A mapping onto a column problem
shows that these systems are zero-range processes, whose steady states
properties are exactly calculated under the assumption of independent column
heights in the
Master equation. Simulation provides accurate island size distributions which
confirm analytic reductions and are particularly useful whenever the analytical
tools cannot provide results in closed form. The shape of
island size distributions can be changed from monomodal to monotonically
decreasing by tuning the temperature or changing the coverage
(one-dimensional density $\rho$). In all cases, small values of the scaling
variable $X\equiv\epsilon^{-1}\rho/\left( 1-\rho\right)$ favour the
monotonically decreasing ones. However, for
large $X$, rapidly increasing rates $\gamma (m)$ lead to  distributions with
peaks very close to $\langle m\rangle$ and
rapidly decreasing tails, while slowly increasing $\gamma (m)$ provide peaks
close to $\langle
m\rangle/2$ and fat right tails.
\end{abstract}

\maketitle

\section{Introduction}

In order to understand the main processes that take place during the growth of
thin films and multilayers, it is important to study the initial steps of those
processes, i. e. the submonolayer regime. The current belief is that insight
derived from the combination of experiment and modeling can lead to better
control of nanostructures formed during deposition. Consequently, models of
submonolayer growth and coarsening with several types of interactions between
the adparticles were intensively studied in the last decades - for a recent
review with applications to homoepitaxial growth, see Ref. \protect\cite{etb}.
Non-equilibrium statistical models are particularly useful for modeling real
systems that are confined to long-life metastable states while some
atomic processes take place in next-to-equilibrium conditions with the
environment.

The recent advances on the production of elongated structures (e. g. nanowires)
along step edges of vicinal surfaces
\cite{Himpsel2001,Gambardella2000,Gai2002,albao,gambardella2006}
and other highly anisotropic growth processes
motivated the study of models that produce effectively one-dimensional
structures on surfaces under certain conditions
\cite{mo,mazzitello,li,ferrando,bentaylor}. However, in many cases it
is advantageous to
restrict the processes of particle diffusion and attachment/detachment from
islands (possibly competing with the continuous atom deposition) to one spatial
dimension, since it enables a more detailed study of the effects of different
physico-chemical parameters.
One of the interesting problems in this field is to determine the island size
distributions under different growth conditions. For instance, recent works
have debated the mechanisms responsible for the onset of monomodal or
monotonically decreasing distributions in submonolayer growth on vicinal
surfaces, since both types of distribution
were already observed experimentally \cite{albao,gambardella2006}.

The peaked island size distributions are the most frequently observed in
growth on planar surfaces \cite{etb} and also appear in some effectively
one-dimensional systems \cite{Gambardella2000,gambardella2006,repain}.
In theoretical models, this feature is observed when some critical
island size is assumed, above which there is no further atom detachment. This
occurs both for deposition competing with diffusion and for post-deposition
coarsening. On the other hand, statistical equilibrium models, where
reversibility of attachment and detachment from islands is implicitly assumed,
provide monotonically decreasing size distributions in one dimension (more
precisely, exponentially decreasing ones) \cite{gambardella2006,tokar}.

The simplest reversible non-equilibrium models consider the processes
illustrated in Fig. 1a in its one-dimensional version (after deposition has
stopped, i. e. with conserved mass) \cite{cv,brune,alanissila,ratch,biehl}.
They account for particle diffusion, aggregation and detachment from islands,
with no critical island size. For simplicity, it is assumed that
attachment occurs immediately upon contact of a diffusing particle with a
cluster, while the detachment occurs with rate $\gamma (m)$, where $m$ is the
mass of the cluster from which the particle leaves. Previous work in
one-dimension only considered the case of constant $\gamma$ for all islands
($m\leq 2$) \cite{coarsen1}, where the steady state presents monotonically
decreasing cluster size distributions, similarly to the equilibrium models 
\cite{gambardella2006,tokar}. The great applicability of the two-dimensional
versions of these models explains the small number of studies of the
one-dimensional cases. 

In this paper, we will consider this class of non-equilibrium one-dimensional
models with detachment rates $\gamma (m)$ increasing with the mass $m$ of the
cluster where the particle is attached. Such dependence is certainly expected
in heteroepitaxial growth if small lattice mismatches between the
adlayer and the substrate are unfavourable to aggregation of new particles to
existing islands. Indeed, the drastic consequences of this feature on island
shapes was already illustrated in some systems, such as $Cu$ islands on
$Ni(100)$, where a
transition from compact to ramified shapes takes place as the coverage
increases \cite{compact_ramified}. In one-dimensional systems, lattice
mismatch is also expected to play a role, although at first approximation the
shape effects are not present and the island mass is sufficient to
determine the rate for detachment of a bordering atom. The association of a
detachment rate with the island size may also be viewed as an alternative to
describe the effects of long-range repulsive interactions present in
a large number of real systems \cite{bogicevic,ovesson,venablesbrune,nandipati},
but which are usually very hard for computation.

The models considered here are equivalent to zero-range processes and do have a
factorisable steady state \cite{evanshanney,spitzer}. This feature means
that the exact calculation of the steady state properties may be carried out
with the use of an independent interval "approximation" (IIA) to the master
equation, which implicitly assumes a factorization of probabilities. The IIA,
which has already proved to be an exceptional tool to investigate
nonequilibrium statistical models \cite{coarsen1,majumdar}, here
provides exact results for some forms of $\gamma(m)$, which are confirmed
by numerical simulation data. However, in many cases the latter approach is
essential to calculate steady state properties, particularly when average
cluster sizes are small and discretization effects play an important role.

We will show that both monomodal and monotonically decreasing cluster size
distributions may be obtained in these models depending on the particular form
of detachment rate and the coverage. In the analysis of some forms of
$\gamma(m)$ which account for short-range attraction of neighboring atoms (via
detachment rates much smaller than isolated particle diffusion coefficients),
we will show that it is possible to exchange between those shapes by tuning the
temperature or changing the coverage - high
temperature and low coverage typically favoring the monotonically decreasing
form. Although no quantitative comparison with real systems data will be shown
here, we believe that the relative simplicity of our model and the range of
qualitative behaviors obtained from it can motivate its use in particular
applications. From the theoretical point of view, this work opens the
possibility of new applications of the widely studied zero-range processes.

The rest of this work is organized as follows. In Sec. II we present the column
picture in which the original problem (Fig. 1) is mapped, the general form of
the master equation in the IIA and the method of solution in the steady state.
In Sec. III we present the steady state cluster size distributions for selected
forms of detachment rates and compare them with results of numerical
simulations. We focus on the differences between limiting cases of rapidly and
slowly increasing $\gamma(m)$. In Sec. IV, we consider systems where formation
of small islands is favoured by detachment rates significantly smaller than
isolated atom ($m=1$) diffusion, while $\gamma(m)$ is increasing. This mimics
the competition between short-range attractive and long-range repulsive
interactions, a case which may be especially relevant for surface science. In
Sec. V we summarize our results and conclusions and discuss possible
applications.

\section{Processes, pictures, and general formulation}

As shown in Fig. 1a, the main processes in our problem are: (i) random walk of
separated (single) particle, at a rate $\gamma(1)$; (ii) detachment of particle
from edge of cluster by particle stepping one unit away from it, at a rate
$\gamma(m)$, where $m$ is the cluster mass. In addition, there is (iii)
attachment of particle to edge of cluster, occurring immediately after the
particle jumps to that position.

The model defined in this so-called cluster picture can also be depicted in a
column picture, as illustrated in Fig. 1b. In the latter, the clusters are
columns, and the group of $n$ ($\geq 1$) vacant sites between two adjacent
clusters has become a group of $n-1$ ($\geq 0$) empty columns between two
filled columns. Thus, one of the vacancies of the cluster picture now acts as a
column spacer, and the remaining ones have become columns with $m=0$ particles. 

This column picture corresponds to a zero-range process \cite{evanshanney}.
Indeed the mapping just demonstrated achieves the equivalent of the inverse of a
mapping from a zero-range process to an exclusion process by other means
\cite{evanshanney,kanai}. In traffic flow models, such as that in Ref.
\protect\cite{kaupuzs}, different rates for jumps to the left and to the right
must be considered.

The full analytic description of systems with such stochastic processes is
provided by the Master equation, which is most easily written in the column
picture. The description is simplified by the fact that the process conserves
the total particle numbers $N$. Thus, using periodic boundary conditions and a
total number of sites $L$ (lattice length in the cluster picture), and denoting
by $N(m)$ the total number of clusters of size $m$ ($\geq 1$), it follows that
(i) $N= \sum_{m=1}^{\infty}{mN(m)}$, (ii) the number of spacers is
$\sum_{m=1}^{\infty}{N(m)}$, and (iii) $N(m)$ equals the number of columns of
size $m$, for $m>0$. Hence, denoting by $N(0)$ the number of columns of size
zero, we have $\sum_{m=0}^{\infty}{N(m)} = L-N\equiv L(1-\rho)$ (the last
step defining the density $\rho$ in the original picture). Thus the total
number of columns (including those of size zero) is constant, as is the
density. The system configuration can be specified by the ordered set of 
numbers of particles in each of the columns in succession: $\left( m_1,m_2
\dots\right) = \{ m_i\}$.

The probability $P_t\{ m_i\}$ at time $t$ of the configuration
$\{ m_i\}$ changes by in and out processes. For example, Fig. 1c shows
the process $\left( m_1,m_2,m_3,\dots\right) \to
\left( m_1+1,m_2-1,m_3,\dots\right)$ ($m_2\geq 1$) having rate $\gamma(m_2)$.
Collecting the effects of all such processes in a time step $t\to t+1$ gives
the full Master equation
\begin{eqnarray}
P_{t+1} \{ m_i\} - P_t \{ m_i\} &=&
\sum_{l=1}^L
[ \gamma\left( m_{l-1}+1\right)
P_t \left( \dots m_{l-1}+1 , m_l-1 \dots\right) \nonumber\\
&& + \gamma\left( m_{l+1}+1\right)
P_t \left( \dots m_{l}-1 , m_{l+1}+1 \dots\right) \nonumber\\
&& - 2 \gamma\left( m_l\right)
P_t \{ m_i\} ]  \theta\left( m_l\right) .
\label{mastergeral}
\end{eqnarray}
The theta function above (zero for $m\leq 0$, otherwise unity) is actually
redundant as $P\left( \dots m-1 \dots\right)$ and $\gamma(m)$ vanish for $m\leq
0$.

As an ansatz, one can attempt to find a solution of the Master equation using
for $P_t\{ m_l\}$ the factorised form
$\prod_{l=1}^L{P_{tl}\left( m_l\right)}$. This turns out to give an approximate
form (the IIA) for the time-dependent situation, but an exact
result for the steady state (the time evolution may be particularly interesting
in the case of rates decreasing with cluster mass, and is the subject of our
current work). In the special case of homogeneous rates ($\gamma$
independent of column position $l$) the steady state Master equation is solved
exactly
with the function $P_{tl}\left( m\right) = P(m)$, i. e. independent of $t$ and
$l$. Here, $P(m)$ is the probability that an arbitrarily chosen column has
occupancy $m$, i. e. $N\left( m\right)/\sum_{m=0}^\infty{N\left( m\right)}$,
which corresponds to the probability that a cluster has mass $m$ in the original
problem (Fig. 1a).

The reduced form of the steady state Master equation applying in this simplified
(homogeneous) situation can be obtained from Eq. (\ref{mastergeral}) or directly
as follows from the processes involved in the evolution of $P(m)$. In
an appropriately defined time step, the change of $P(m)$ has positive and
negative contributions (from in and out processes)
\begin{equation}
P(m+1) \gamma(m+1) \theta(m) + \delta_{m,0} \gamma(1) P(1) + \Gamma
\left[ P(m-1) \theta(m-1) + \delta_{m,1} P(0) \right]
\label{in}
\end{equation}
and
\begin{equation}
-P(m) \gamma(m) \theta(m-1) - \delta_{m,1} \gamma(1) P(1) - \Gamma \left[ P(m)
\theta(m) + \delta_{m,0} P(0) \right] ,
\label{out}
\end{equation}
respectively, where
\begin{equation}
\Gamma \equiv \sum_{l=1}^\infty{\gamma(l)P(l)} .
\label{defGamma}
\end{equation}

The Master equation for the steady state now reduces to the condition (on
$P(m)$) that the sum of contributions of in and out processes (Eqs. \ref{in} and
\ref{out}) vanishes for each $m$. Defining
\begin{equation}
A(m) \equiv P(m)\gamma(m) - \Gamma P(m-1) , m\geq 1 ,
\label{defA}
\end{equation}
that gives the exact relations
\begin{eqnarray}
A(1) &=& 0 ,  \nonumber\\
A(m+1) - A(m) &=& 0 \qquad ,\qquad m\geq 1 ,
\label{eqA}
\end{eqnarray}
so that $A(m)$ vanishes for all $m\geq 1$. Thus $P(m)$ satisfies
\begin{equation}
P(m) = \frac{\Gamma}{\gamma\left( m\right)} P(m-1) , m\geq 1 ,
\label{Pestac1}
\end{equation}
yielding
\begin{equation}
P(m) = P(0)\prod_{l=1}^m{\frac{\Gamma}{\gamma\left( l\right)}} , m\geq 1 .
\label{Pestac2}
\end{equation}
Direct substitution of the resulting product form for $P\{m\}$ into the steady
state version of the original Master equation (\ref{mastergeral}) verifies that
this satisfies it exactly. This result is equivalent to the one \cite{spitzer}
previously given for zero-range processes (see e. g. Ref.
\protect\cite{evanshanney}).

Using the normalization condition
\begin{equation}
\sum_{m=0}^\infty{P\left( m\right)} = 1 ,
\label{normalisation}
\end{equation}
it can be seen that Eq. (\ref{Pestac2}) is consistent with the definition
(\ref{defGamma}).

Using the definitions of Sec. II, we have
\begin{equation}
{\langle m\rangle}_{all} \equiv
\frac{\sum_{m=0}^\infty{mN\left( m\right)}}{\sum_{m=0}^\infty{N\left( m\right)}}
= 
\sum_{m=0}^\infty{mP\left( m\right)} = \frac{N}{L\left( 1-\rho\right)} =
\frac{\rho}{1-\rho},
\label{mmediogeral}
\end{equation}
where ${\langle m\rangle}_{all}$ is the mean cluster size taking into account
all columns, including those with zero mass. These equations provide a means to
relate $P(0)$ in Eq. (\ref{Pestac2}) to the density $\rho$, as well as the
relationship between ${\langle m\rangle}_{all}$ and $\rho$.

Defining $\mu\equiv \ln{\Gamma}$ and
\begin{eqnarray}
s(m) &\equiv & \sum_{l=1}^m{\ln{\gamma\left( l\right)}} , m\geq 1 , \nonumber\\
s(0) &\equiv & 0 , m=0 ,
\label{defs}
\end{eqnarray}  
and using Eqs. (\ref{Pestac2}) and (\ref{normalisation}), we find
\begin{equation}
P(m) = \exp{\left[ m\mu - s\left( m\right)\right]} / 
\sum_{l=0}^\infty{\exp{\left[ l\mu - s\left( l\right)\right]}} .
\label{Psolucao}
\end{equation}
From Eq. (\ref{mmediogeral}), $\mu$ can be determined in terms of $\rho$ by
\begin{equation}
\frac{\rho}{1-\rho} =
\sum_{m=0}^\infty{m\exp{\left[ m\mu - s\left( m\right)\right]}} /
\sum_{m=0}^\infty{\exp{\left[ m\mu - s\left( m\right)\right]}} .
\label{rhomu}
\end{equation}
The limits in the sums are those appropriate for an infinite system ($N , L \to
\infty$ at fixed density $\rho=N/L$). The sums are constrained for finite $L$,
$N$.

For the infinite system there remains the question whether the sums converge or
not. That depends on the form of the rates $\gamma(m)$ at large $m$.
In that region $m$ can be treated as a continuous variable and sums become
integrals. This continuum approach is also very useful for $\rho$ near $1$,
where typical representative $m$'s (such as ${\langle m\rangle}_{all}$ in Eq.
\ref{mmediogeral}) are large, so typically $P(m)$ is appreciable at large $m$.
Using such a continuum approach,
it can be seen that for $\gamma(m)$ increasing with $m$ at large $m$ (making
$\ln{\gamma\left( m\right)}$ positive and increasing), $s(m)$ increases with $m$
more rapidly than linearly, making the sums in Eqs. (\ref{normalisation}),
(\ref{Psolucao}) and (\ref{rhomu}) converge, so 
the steady state solution for $P(m)$ is physically acceptable: $P(m)$ decreases
with $m$ at large $m$ and is typically peaked. This is exactly the situation of
interest in real systems where the increase of island size is unfavourable.

In the continuum approach, the location $m=m_0$ of the peak in $P(m)$ can be
found using 
\begin{equation}
0 = {\left[ \frac{d}{dm}
\left[ \mu m-s\left( m\right)\right] \right] }_{m=m_0} = \mu -
\ln{\gamma\left( m_0\right)} .
\label{mmax}
\end{equation}
For ${m_0}^2\gamma'\left( m_0\right)/\gamma\left( m_0\right) \gg 1$ the peaking
is strong, and $P(m)$ can be approximated by ${\left[ \frac{\gamma
\left( m_0\right)}{2\pi\gamma'\left( m_0\right)} \right]}^{1/2}
\exp{\left[ \mu\left( m-m_0\right) - I\left( m\right) \right]}$, where
$I\left( m\right) \equiv \int_{m_0}^m{\ln{\left[ \gamma\left( l\right)\right]}
dl}$. This has most of its weight in the Gaussian
form ($P(m)\propto \exp{\left[ -\frac{1}{2} \left( \gamma'\left( m_0\right) /
\gamma\left( m_0\right) \right) {\left( m-m_0\right)}^2\right]}$), which applies
near the peak.

For $\gamma(m)$ decreasing with $m$ sufficiently fast at large $m$, the sums
diverge and there is strictly  no steady state in the infinite system. It can be
shown that the criterion for no steady state in the infinite system is
$\gamma(m)$ decaying at large $m$ more slowly than a constant times
$1+\frac{2}{m}$ \cite{evanshanney}. In this case, the system will coarsen
forever, but such situations [$\gamma \left( m\right)$ decreasing with $m$]
will not be considered here.

\section{Steady state behavior for selected detachment rates}

In this Section, analytic predictions and simulation results for steady state
properties will be presented and compared. The emphasis will be on the
empirically realistic "thermodynamic" limit of very large systems, in which
steady states are achievable.
The steady state properties to be discussed here are the cluster size
distributions and the average cluster sizes. However, it is important to
mention that, from now on, cluster sizes are defined by averaging only over
masses $m\geq 1$, in contrast with Eq. (\ref{mmediogeral}), which also took
into account columns with zero mass. This new average will be denoted ${\langle
m\rangle}$, and provides a more appropriate physical description of the system
in the original cluster picture. Indeed, except where explicitly indicated, we
will refer to that original picture in the following.

The simulations were typically performed in lattices of sizes $L=8192$, with
several densities and different forms of $\gamma(m)$. Due to the small
cluster sizes imposed by the system dynamics, in all cases finite-size effects
are negligible (this was checked by comparison of results in different lattice
sizes). The generation of a sequence of configurations begins with the
deposition of a random layer of density $\rho$. Subsequently, the dynamics with
diffusion, attachment and detachment processes is allowed, and the evolution of
the cluster size distribution is monitored. Typically, it is assumed that the
steady state is attained if no appreciable change (e. g. $1\%$ in the peak) is
found in the distribution at the last time decade of the simulation, after
averaging over at least 100 initial configurations (i. e. 100 different
sequences). Under these conditions, the 
representative configurations are attained, on the average, after $100-1000$
detachments of particles aggregated to small clusters. Anyway,
it is important to note that in our models steady state properties are
independent of the initial system
configuration, thus the same results would be obtained if diffusion and
detachment processes were competing with deposition during the production of
the first configuration with the desired density $\rho$ (however, there are
special models where steady state properties depend on the initial
configuration - see e. g. Ref. \protect\cite{grynberg}).

First we consider the case (i) of constant detachment rate, $\gamma(m)=a$ for
$m\geq 2$ and isolated particle hopping rate $\gamma(1) = b$. The analysis
proceeding from Eqs. (\ref{rhomu}) and
(\ref{Psolucao}) gives a size distribution which is exponentially decreasing in
$m$ for $m\geq 2$. For the special case $b=a$, using
$s(m)=m\ln{a}$, we have $P(m)=\left( 1-\rho\right)\rho^m$ for all
$m$. Such results were formerly predicted in Ref. \protect\cite{coarsen1} and
numerically confirmed in Ref. \protect\cite{chamereis}. These distributions are
equivalent to those obtained in equilibrium (reversible) models with nearest
neighbor interactions between the adatoms \cite{gambardella2006,tokar}. 

Now we consider the power law case (ii), $\gamma(m) = m^k$ for $m\geq 1$, in
which the reduction of the analytic result (\ref{Psolucao}) for $P(m)$ is less
straightforward than in the previous case. For this reason, most results are
obtained from simulation. In Fig. 2, we show the scaled distributions
for $k=1$ and densities $\rho=0.5$, $\rho=0.75$ and $\rho=0.95$ (in this and
subsequent plots, dashed curves are guides to the eye). For small and
medium densities, isolated particles ($m=1$) are predominant
and the distribution is rapidly decreasing,
so that it seems to decay faster than a simple exponential for
$\rho\lesssim 0.5$. As the coverage increases, it crosses over to a peaked
(monomodal) distribution, which becomes very sharp for $\rho$ close to $1$. This
crossover is directly related to the increase in the average size ${\langle
m\rangle}\approx \rho/\left( 1-\rho\right)$, from a value near unity for small
$\rho$ (where the decrease is monotonic) to large values for large $\rho$
(where the peak is close to ${\langle m\rangle}$).
 
The analytical distribution of case (ii) becomes simple for $\rho\approx 1$,
since there the
typical masses are large and continuum approximations can be used. That results
in $P(m) \propto \exp{\left[ \left( \mu+k\right)m - km\ln{m}\right]}$ ($m$
large). For $k=1$, this is the large $m$ approximation to a Poisson
distribution. The agreement with simulation results is illustrated in the inset
of Fig. 2 for $k=1$ and $\rho=0.95$, where
$\log{\left[ P\left( m\right)\right]}$ is shown to be a function of
$m\ln{\left( m\right)}-Cm$, with a fitting constant $C=3.92$.

We next consider another simple detachment rate function, $\gamma(m) =
{\left( \frac{m}{m+1}\right)}^b$ for $m\geq 1$, hereafter called case (iii). The
main difference
from case (ii) is the fact that $\gamma$ does not diverge as $m\to\infty$.
Using Eqs. (\ref{rhomu}) and (\ref{Psolucao}), we obtain exactly $P(m) \propto
{\left( m+1\right)}^b \exp{\left( -\beta m\right)}$, with $\beta$ constant.
For low integer values of $b$, the constant $\beta$ and the normalization
constant can easily be obtained analytically in terms of the density. For the
simplest case $b=1$, we obtain $\beta = \ln{\left( \frac{2-\rho}{\rho}\right)}$
and $\langle m\rangle = \frac{{\left( 2-\rho\right)}^2}{\left( 1-\rho\right)
\left( 4-3\rho\right)}$.
We remark here that the case $b=-2$ is marginal: from the discussion at the end
of Sec. II (see also Ref. \protect\cite{evanshanney}), the infinite
system achieves a steady state if $b\geq -2$ but not if $b<-2$.

In Fig. 3 we show the scaled cluster size distributions for case (iii) obtained
from simulation, with $b=1$ and coverages $\rho=0.5$ and $\rho=0.9$. Both show
excellent agreement with the analytical results (hereafter represented by solid
curves in the plots). Again we observe that the distribution is
monotonically decreasing for small coverages, where $\langle m\rangle$ is close
to $1$, while for larger coverages there appears a
peak. However, there is an important difference from case (ii) here: instead of
having a sharp peak close to  $\langle m\rangle$, the distribution for large
coverages has a very fat left tail and the most probable island size may be
smaller than $\langle m\rangle/2$ (see data for $\rho=0.9$ in Fig. 3). This
difference is certainly a consequence of the slower increase of the detachment
rate with the cluster mass.

We also analyzed other forms of detachment rates, namely simple exponential,
logarithmic $\gamma(m)$, and the particular logarithmic case
$\gamma(m)=c{ \left[ \ln{\left( m+1\right)}
\right] }^{\left[ 1+\ln{\left( m+1\right)}\right]}$, for which a closed
analytical form of $P(m)$ can be obtained. In all cases, the analytically
predicted distributions agree very well with simulation data. With the
logarithmic forms, the qualitative behavior is intermediate between cases (ii)
and (iii) above.

The peaked cluster size distributions shown above were always obtained for large
densities. At first sight, this condition seems to be unattainable in
experiments on submonolayers on vicinal surfaces, since the one-dimensional
character of the adlayer (e. g. chain-like structures along the step edges)
is lost at high coverages. However, in Sec. IV, we will show that under certain
realistic conditions on $\gamma(m)$, it is also possible to observe peaked
distributions for densities lower than $0.5$.

Moreover, it is important to stress that the density $\rho$ of particles along
the steps can be much larger than the nominal coverage $\theta$ of the surface
because the latter is an adatom density per substrate area, while $\rho$ is
the filling of one-dimensional rows. As an example, we refer to the STM image
of $Ag$ deposited on $Pt(997)$ in Fig. 1b of Ref.
\protect\cite{gambardella2006}), where the coverage $\theta=0.04$ monolayers is
much smaller than the effective filling $\rho\approx 0.3$ of the step rows. 

\section{Systems with competing interactions and temperature effects}

Here we extend the study of Sec. III to systems with the mass-dependent
detachment rates all far less than the isolated atom diffusion
rate $\gamma(1)$, by an overall factor $\epsilon\sim \exp{\left( -E/k_B
T\right)}\ll 1$. Here, $E$ may be interpreted as a binding energy between
neighboring adatoms, and $T$ is the temperature (see e. g. the discussion on the
diffusivities for initial state interactions in  Ref. \protect\cite{payne},
where only short-range interactions were considered). For low temperatures,
that factor significantly reduces the mobility of aggregated
atoms when compared to the isolated ones. On the other hand, the mass dependence
of $\gamma$ accounts for the effects of the interactions with the substrate,
which works against the increase of cluster size (e. g. effects of lattice
mismatch or substrate-mediated repulsive interactions). We will typically work
with ${10}^{-1}\leq\epsilon\leq {10}^{-3}$ for each form of $\gamma(m)$, in
order to show the possible effects of the temperature in the island size
distributions. Simulation work here will focus on low coverages, typically below
or at half filling of the one-dimensional rows.

First we consider the generalization of case (ii), where
$\gamma{\left( m\right)} = c\left( m\right) m^k$ with $c(1)=1$, and
$c(m)=\epsilon\ll 1$ for $m\geq 2$. This means that detachment rates from small
clusters (up to masses $\approx \epsilon^{-1}$) are smaller than the isolated
atom diffusion rate, but larger clusters are very unstable.

For $k=1$, no simple closed form for the cluster size distribution can be
obtained. In Figs. 4a and 4b we show the simulation results with densities
$\rho=0.1$ and $\rho=0.5$, respectively. In both plots we consider
$\epsilon={10}^{-2}$ and $\epsilon={10}^{-3}$. For low
density (Fig. 4a), the distributions are monotonically decreasing up to
$\epsilon\sim {10}^{-2}$, but a peak at small $m$ appears at sufficiently low
temperatures (i. e. very small $\epsilon$). For medium density, Fig. 4b shows
that the temperature does not need to be so small for the onset of a peaked
distribution: the peak is present for $\epsilon={10}^{-2}$ and it is well
defined at $\epsilon={10}^{-3}$. These results must be compared with those in
Fig. 2 for $k=1$ and $\rho=0.5$, but $\epsilon =1$, where the distribution
is monotonically and rapidly decreasing.

Analytical results can be obtained for the generalized case (ii) only for $k<1$
and $\epsilon \ll 1$, where average cluster sizes are large and a continuum
approximation of the cluster size distribution is possible. In these
conditions, we have $P\left( m\right) \propto \frac{1}{{\left( m!\right)}^k}
{\{ \frac{1}{k} \ln{\left[ B/{\left( \ln{B}\right)}^\zeta \right]}
\}}^{km}$, with $B\equiv \frac{k^{2-k/2}}
{{\left( 2\pi\right)}^{\left( 1-k\right)/2}}
\frac{\rho}{\epsilon\left( 1-\rho\right)}$ and $\zeta = \frac{3-k}{2}$. This is
a monomodal
distribution with $\langle m\rangle \approx \frac{1}{k}
\ln{\left[ B/{\left( \ln{B}\right)}^\zeta \right]}$, which confirms the
general trend that
this shape is favoured by large densities and small temperature (small
$\epsilon$).

Unfortunately, the above formula for $P(m)$ is accurate only for very small
$\epsilon$ and very small $k$, otherwise the continuum approximation fails due
to the discreteness of the typical cluster sizes and the large statistical
weight of isolated particles. As an example, we compare in Fig. 5a and 5b the
analytical and numerical cluster size distributions for $k=1/2$ and $\rho
=0.5$, with $\epsilon = {10}^{-2}$ and $\epsilon={10}^{-3}$, respectively. Even
for  $\epsilon={10}^{-3}$, where $\langle m\rangle \approx 7$, we observe
deviations of the analytical approximation from the numerical data.

The results in Figs. 4 and 5, as well as the analytical approximation for $k<1$
and $\epsilon\ll 1$, show another important feature: the peaks of the
distributions are  very close to $\langle m\rangle$, similarly to the case (ii)
with $\epsilon =1$ studied in Sec. III. We recall that these are cases of
rapidly increasing $\gamma(m)$, in which the formation of large clusters is
highly unfavourable.

Now we consider the generalization of case (iii), where $\gamma(m) =
c\left( m\right) {\left( \frac{m}{m+1}\right)}^b$, with $c(1)=1$, and
$c(m)=\epsilon$ for $m\geq 2$. The resulting probability distribution
$P(m)$ is of the same form as the one in Sec. III (i. e. for the special case
$\epsilon =1$), however the small $\epsilon$ gives much lower weight to $P(m)$,
$m\geq 1$, than $P(0)$, and the peak of the distribution is, for most
densities, pushed out to much larger masses. For general $b$ it is possible to
show that for small enough $\epsilon$ (very much less than both $\rho$ and
${\left( 1-\rho\right)}^{b+1}$) the mean cluster size has the form ${\langle
m\rangle} \sim {\left[ \frac{\rho}{\epsilon\left( 1-\rho\right)}
\frac{{\left( b+1\right)}^{b+1}}{b!}\right]}^{1/\left( b+2\right)}$.

In Figs. 6a-d we show the simulation results for $b=1$, with densities
$\rho=0.1$ and $\rho=0.5$ and detachment factors
$\epsilon={10}^{-2}$ and $\epsilon={10}^{-3}$. For this value of $b$, the full
distribution can be obtained analytically as $P(m) \propto \left( m+1\right)
e^{-\beta m}$, $\beta = {\left( \frac{2\epsilon\left( 1-\rho\right)}{\rho}
\right)}^{1/3}$. It compares well with simulation data for intermediate
densities and small values of $\epsilon$, as shown in Fig. 6b.

Lowering the
temperature also provides peaked distributions in this case, and the position of
the peak is shifted to larger masses as the density increases. The effect of
temperature is important only in the right tail of the distribution, which
typically has a small number of data points, corresponding to small islands.
Similarly to what was observed in Sec. III, the peaks are located close to
$\langle m\rangle /2$ and the left tail is very fat. This is a signature of a
weak size-dependence of the detachment rates. However, the main difference from
the results in Sec. III is again the possibility of finding the transition from
a monotonically decreasing distribution to a monomodal one by tuning the
temperature, with a coverage not too large.

For the general case of rate functions $\gamma (m)$ increasing with $m$ (at
large $m$), the mean cluster size ${\langle m\rangle}$ can always be made
large at low densities by suppressing detachment rates by a suitably small
factor $\epsilon$, compared to the diffusion rate $\gamma(1)$. Considering
$\gamma(m) = \epsilon\nu(m)$ for $m\geq 2$ and $\gamma(1)=\nu(1)$,
the relationships
\begin{equation}
{\langle m\rangle} \sim F(X) \qquad ,\qquad X \equiv 
\frac{\rho}{\epsilon\left(1-\rho\right)}
\label{defX}
\end{equation}
of average cluster size to density and Arrhenius rate parameter $\epsilon$ which
result at small $\epsilon$ for a variety of detachment rate functions (at
large $m$) are:
\begin{eqnarray}
\nu(m) = m^k &\Rightarrow & F(X) = \frac{1}{k}\ln{X} ,\nonumber\\
\nu(m) = {\left(\ln{m}\right)}^{\alpha\ln{m}} &\Rightarrow & F(X) =
\frac{\ln{X}}{\alpha\ln{\ln{\ln{X}}}} ,\nonumber\\
\nu(m) = {\left[ me^{{\left( \ln{m}\right)}^2}\right]}^\alpha &\Rightarrow &
F(X) = \frac{\ln{X}}{\alpha\ln{\ln{X}}} ,\nonumber\\
\nu(m) = e^{\alpha m^n} &\Rightarrow & F(X) =
{\left[ {\frac{n+1}{\alpha n} \ln{X}} \right]}^{1/\left( n+1\right)} .
\label{listaF}
\end{eqnarray}
It can be seen that the effect of $\epsilon$ (in reducing the density required
to get a large mean cluster mass) diminishes as the rate of increase of $\nu$
with $m$ increases.  More importantly, these results show that density and
temperature control the island size distribution through the scaling variable
$X$, so that monomodal (monotonically decreasing) forms are found for large
(small) $X$.

\section{Discussion and Conclusion}

We studied one-dimensional models of island formation by diffusion, attachment
and detachment of single particles, considering detachment rates $\gamma$
increasing with the island mass $m$. This type of model may be of experimental
interest because it can be used to simplify the description of more complex
interactions that prevent the formation of large atomic islands. Indeed, these
models are equivalent to zero-range processes in which the system attains a
steady state. An independent interval "approximation" to the Master equation
was proposed in order to calculate the cluster size distribution which actually
provides an exact description in the steady state. The tools necessary for
derivation of explicit distributions for any particular form of $\gamma(m)$
were provided and results for a variety of cases were presented
and compared with numerical simulation data.

The representative rate functions analyzed above include
some which arise from associating (Arrhenius) detachment rates with potentials
$U(m)$ for particles at the end of a cluster of size $m$. $U(m)$ is then a sum,
from $l=1$ to $m-1$, of pair potentials $V(l)$ for separation $l$. The forms
$V$ vanishing, $V$ Coulomb-like and inverse square then give
(exactly in the first two cases, and for large $m$ in the third one) the
detachment rate functions of cases (i), (ii), (iii), respectively. In addition,
we also mimicked the presence of short-range attraction in the models where the
detachment rates were smaller than isolated particle hopping rates by an
overall factor $\epsilon\ll 1$.

One of the important conclusions of this work is the possibility of changing the
shape of island size distributions, from monomodal to monotonically decreasing
ones, by tuning the temperature or changing the coverage. For different forms
of detachment rates (rapidly or slowly decreasing with $m$), it was shown that
high temperatures and low coverages are favourable for the onset of the
monotonically decreasing distributions, their combined effect being described by
the scaling variable $X\equiv \frac{\rho}{\epsilon\left( 1-\rho\right)}$. 
This transition is not usual in non-equilibrium modeling of submonolayer
growth, particularly when irreversible attachment to islands (i. e. a critical
island size) is assumed. However, it is an important finding because both
shapes of distribution were already observed (isolated) in island growth near
step edges of vicinal surfaces \cite{albao,gambardella2006}. 

On the other hand, rapidly or slowly increasing $\gamma(m)$ have different
effects on the position of the peak of the island size distributions and on its
right tail. Rapidly increasing $\gamma(m)$ [e. g. case (ii) above] lead to
peaks very close to $\langle m\rangle$ and rapidly decreasing tails, since
formation of large clusters is very difficult. However, the slowly decreasing
$\gamma(m)$ forms [e. g. case (iii), where $\gamma(m)$ does not diverge when
$m\to\infty$] provide peaks close to $\langle
m\rangle/2$ and fat right tails. In real systems where island size
distributions are measured, these results may give clues on how intense are the
mechanisms that work against the formation of large clusters.

Despite the fact that no application to a particular real system was proposed
here, we believe that the framework developed in this paper may be useful for
such applications. Long range interactions frequently play a
role in submonolayer growth but introduce difficulties to both analytical
(scaling) and numerical calculations, even when they are limited to one
dimension (see. e. g. \cite{ammi}). Consequently, the inclusion of simpler
mechanisms in a model system may be useful, such as the association of
detachment rates to the full cluster size suggested here.
The fact that our model corresponds to a zero-range process allows
for much simpler analytical calculations of steady state properties, in
contrast to other (not less important) approaches, such as the introduction of
energy barriers for particle attachment to clusters
\cite{kandel,venablesbrune}. Finally, it is also interesting to recall that gas
adsorption in carbon nanotubes may be viewed as a one-dimensional clustering
problem, thus it is another field where this type of non-equilibrium model may
find application. Indeed, simple statistical equilibrium models of interacting
particles in finite lattices were already proposed for those systems
\cite{hodak}.

\begin{acknowledgments}

We wish to thank Dr. Rosemary Harris for a discussion concerning zero-range
process results for finite systems.

FDAA Reis thanks the Rudolf Peierls Centre for Theoretical Physics of Oxford
University, where this work was done, for hospitality, and acknowledges
support by the Royal Society of London (UK) and Academia Brasileira de
Ci\^encias (Brazil) for his visit.

RB Stinchcombe acknowledges support
from the EPSRC under the Oxford Condensed Matter Theory Grants,
numbers GR/R83712/01, GR/M04426 and EP/D050952/1.

\end{acknowledgments}

\newpage

\begin{figure}[!h]
\includegraphics[clip,width=0.5\textwidth,angle=0]{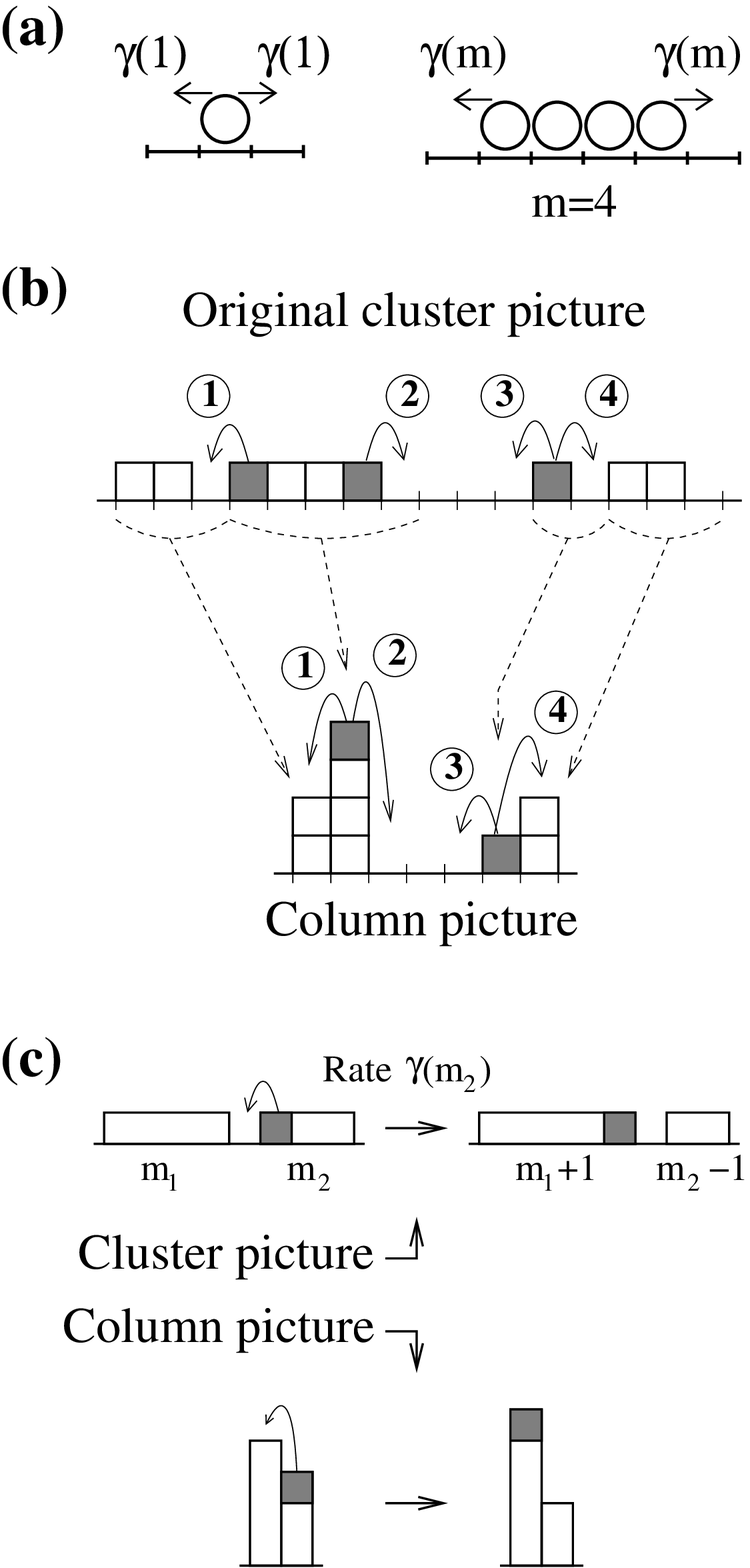}
\caption{(a) Illustration of the diffusion ($m=1$) and detachment
($m>1$)
processes of the model, with the associated rates $\gamma(m)$. (b) Examples of
detachment processes (1,2) and hopping processes (3,4) of filled particles, in
the original cluster picture and in the corresponding column picture. Dashed
lines show the correspondence between clusters in the two pictures. (c) Example
of leftward movement of an aggregated particle.}
\label{fig1}
\end{figure}

\begin{figure}[!h]
\includegraphics[clip,width=0.7\textwidth,angle=0]{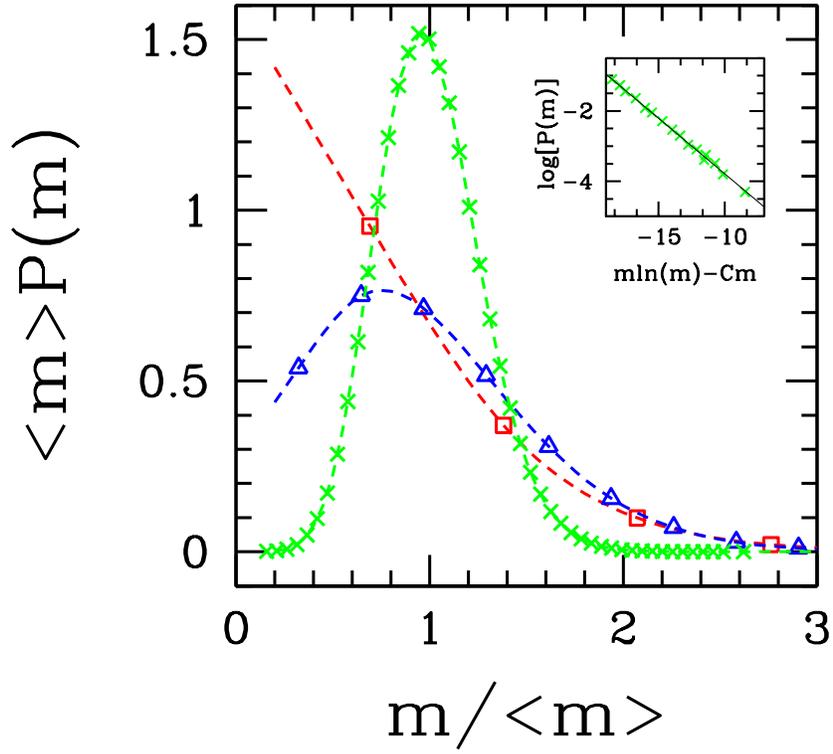}
\caption{(Color online) Scaled cluster mass distribution for detachment rate
$\gamma(m)=m$
(case (ii) with $k=1$), with densities $\rho=0.5$ (squares), $\rho=0.75$
(triangles) and $\rho=0.95$ (crosses). Dashed curves are drawn to guide the eye.
Inset: the rescaled data for $\gamma(m)=m$ and $\rho=0.95$, with a fitting
constant $C=3.92$, is consistent with a Poisson distribution (the solid line
is a linear fit of the data).}
\label{fig2}
\end{figure}

\begin{figure}[!h]
\includegraphics[clip,width=0.7\textwidth,angle=0]{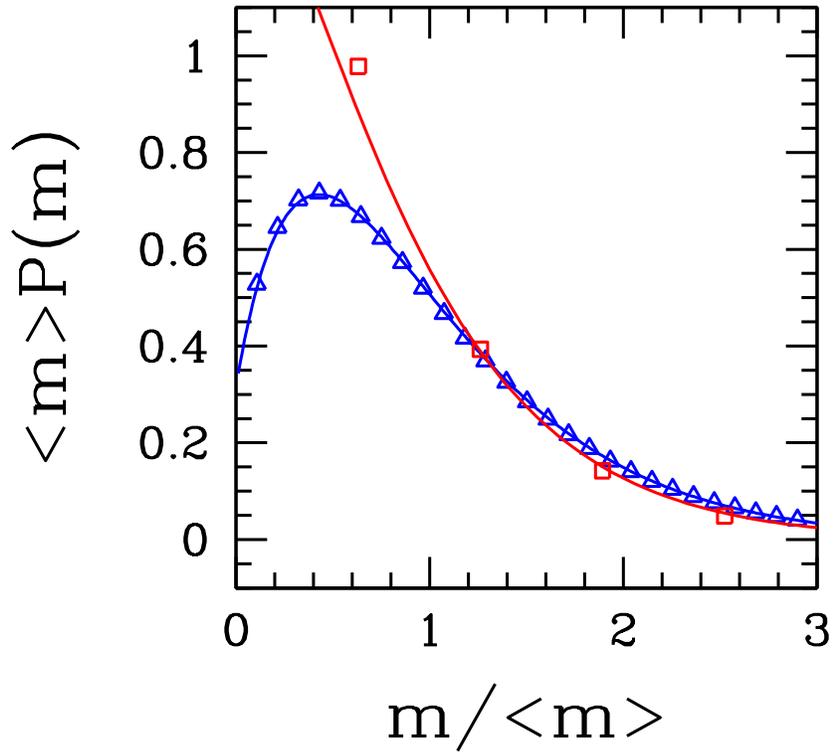}
\caption{(Color online) Scaled cluster mass distribution obtained from
simulation for detachment rate
$\gamma(m)=m/{\left( m+1\right)}$ [case (iii)] with densities $\rho=0.5$
(squares) and $\rho=0.9$ (triangles). Solid curves are the corresponding
analytical results.}
\label{fig3}
\end{figure}

\begin{figure}[!h]
\includegraphics[clip,width=0.7\textwidth,angle=0]{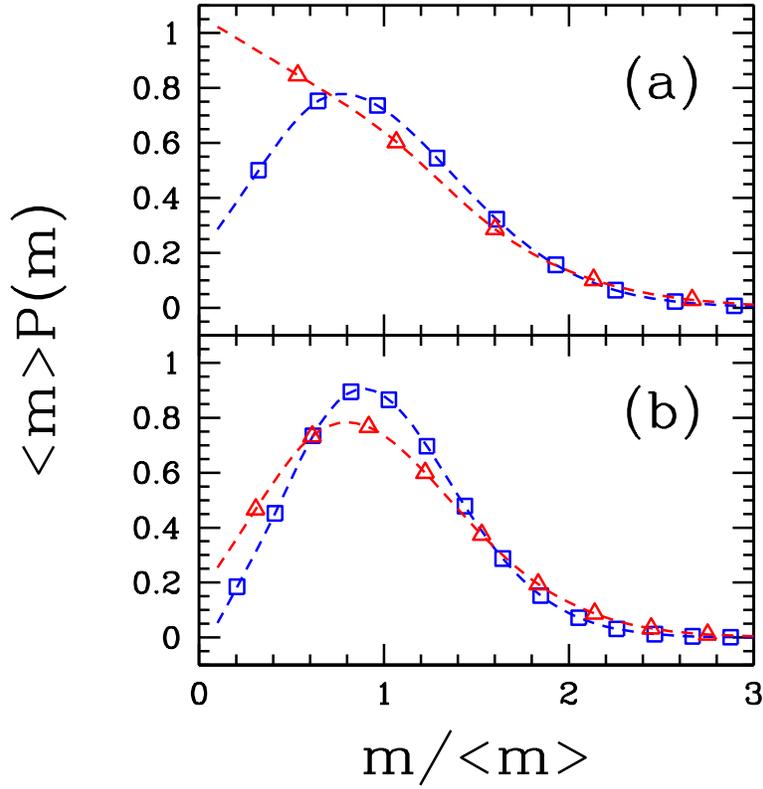}
\caption{(Color online) Scaled cluster mass distributions obtained from
simulation for detachment rate
$\gamma(m)={c\left( m\right)} m$, with $c(1)=1$ and $c(m)=\epsilon$ for $m\geq
2$, and densities (a) $\rho=0.1$ and (b) $\rho=0.5$. In both panels,
$\epsilon={10}^{-2}$ (triangles) and $\epsilon={10}^{-3}$ (squares). Dashed
curves are drawn to guide the eye.}
\label{fig4}
\end{figure}

\begin{figure}[!h]
\includegraphics[clip,width=0.7\textwidth,angle=0]{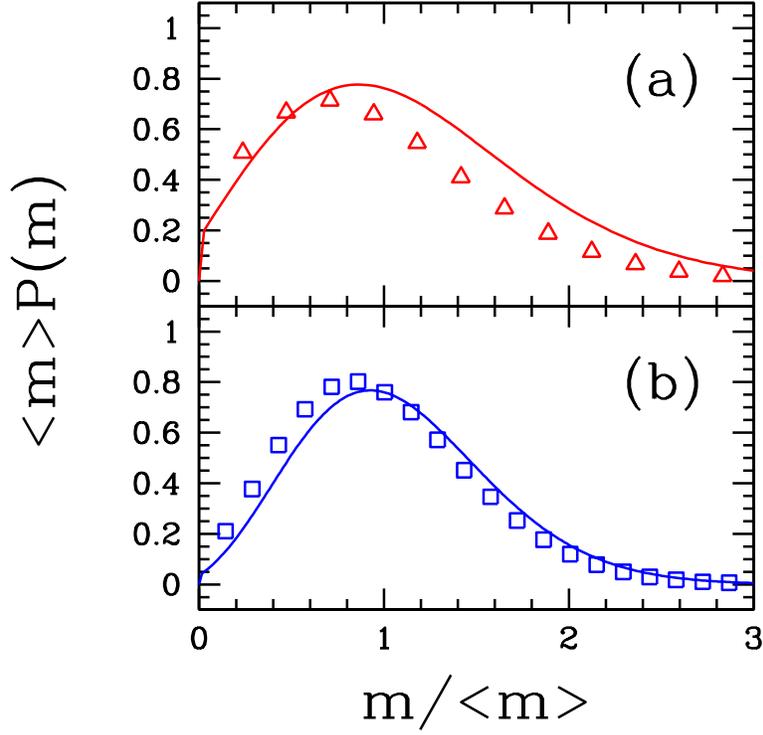}
\caption{(Color online) Scaled cluster mass distributions for detachment rate
$\gamma(m)={c\left( m\right)} m^{1/2}$, with $c(1)=1$ and
$c(m)=\epsilon$ for $m\geq 2$, density $\rho=0.5$ and (a) $\epsilon={10}^{-2}$
and (b) $\epsilon={10}^{-3}$. Symbols indicate simulation data and solid curves
show the corresponding analytical results  in the continuum approximation.}
\label{fig5}
\end{figure}

\begin{figure}[!h]
\includegraphics[clip,width=0.7\textwidth,angle=0]{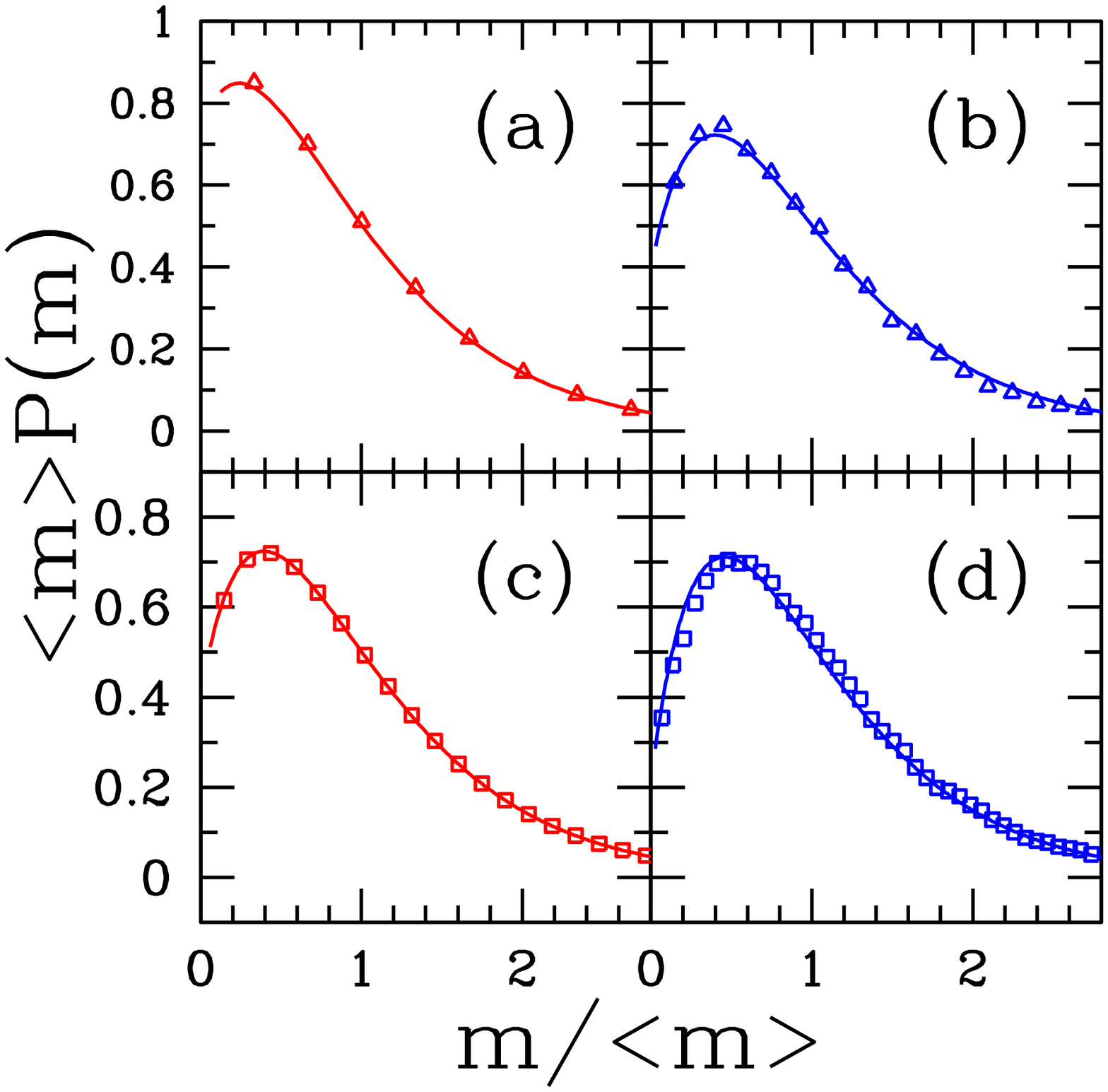}
\caption{(Color online) Scaled cluster mass distributions for detachment rate
$\gamma(m)={c\left( m\right)} m/{\left( m+1\right)}$, with $c(1)=1$ and
$c(m)=\epsilon$ for $m\geq 2$, and: (a) $\rho=0.1$, $\epsilon={10}^{-2}$; (b) 
$\rho=0.1$, $\epsilon={10}^{-3}$; (c) $\rho=0.5$, $\epsilon={10}^{-2}$; (d)
$\rho=0.5$, $\epsilon={10}^{-3}$.  Symbols indicate simulation data and solid
curves show the corresponding analytical results.}
\label{fig6}
\end{figure}


\begin{references}

\bibitem{etb}
J.W. Evans, P. A Thiel, M. C. Bartelt, Surface Science 
Reports {\bf 61}, 1 (2006).

\bibitem{Himpsel2001} F.J. Himpsel, A. Kirakosian , J. N. Crain, J. L. Lin, and
D. Y. Petrovykh, Solid State Communications {\bf 117}, 149 (2001).

\bibitem{Gambardella2000} P. Gambardella, M. Blanc, H. Brune,
K. Kuhnke, and K. Kern, Phys. Rev. B {\bf 61}, 2254 (2000).

\bibitem{Gai2002}  Z. Gai, G. A. Farnan. J. P. Pierce and J. Shen, App.
Phys. Lett. {\bf 81}, 742 (2002).

\bibitem{albao}
M. A. Albao, M. M. R. Evans, J. Nogami, D. Zorn, M. S. Gordon and J. W. Evans,
Phys. Rev. B {\bf 72}, 035426 (2005).

\bibitem{gambardella2006}
P. Gambardella, H. Brune, K. Kern, and V. I. Marchenko, Phys. Rev. B {\bf 73},
245425 (2006).

\bibitem{mo}
Y. W. Mo, J. Kleiner, M. B. Webb, and M. G. Lagally, Phys. Rev. Lett. {\bf 66},
1998 (1991).

\bibitem{mazzitello}
K. I. Mazzitello, J. L. Iguain, and H. O. Martin, J. Phys. A {\bf 32}, 4389
(1999).

\bibitem{li}
Y. Li, M. C. Bartelt, J. W. Evans, N. Waelchli, E. Kampshoff, and K. Kern, Phys.
Rev. B {\bf 56}, 12539 (1997).

\bibitem{ferrando}
R. Ferrando, F. Hontinfinde, and A. C. Levi, Phys. Rev. B {\bf 56}, R4406
(1997).

\bibitem{bentaylor}
J. B. Taylor and P. H. Beton, Phys. Rev. Lett. {\bf 97}, 236102 (2006).

\bibitem{repain}
V. Repain, G. Baudot, H. Ellmer, and S. Rousset, Mater. Sci. Engineering B {\bf
96}, 178 (2002).

\bibitem{tokar}
V. I. Tokar and H. Dreyss\'e, Phys. Rev. B {\bf 74}, 115414 (2006).

\bibitem{cv}
S. Clarke and D. D. Vvedensky, J. Appl. Phys. {\bf 63}, 2272 (1988).

\bibitem{brune}
H. Brune, Surf. Sci. Rep. {\bf 31}, 121 (1998).

\bibitem{alanissila}
T. Ala-Nissila, R. Ferrando, and S. C. Ying, Adv. Phys. {\bf 51}, 949 (2002).

\bibitem{ratch}
C. Ratch and J. A. Venables, J. Vac. Sci. Technol. A {\bf 21}, S96 (2003).

\bibitem{biehl}
M. Biehl, arXiv:cond-mat/0406707 (2004).

\bibitem{coarsen1}
F. D. A. Aar\~ao Reis and R. B. Stinchcombe, Phys. Rev. E {\bf 70}, 036109
(2004).

\bibitem{compact_ramified}
B. M\"uller, L. Nedelmann, B. Fischer, H. Brune, J. V. Barth, and K. Kern, Phys.
Rev. Lett. {\bf 80}, 2642 (1998).

\bibitem{bogicevic}
A. Bogicevic, S. Ovesson, P. Hyldgaard, B. I. Lundqvist, H. Brune and D. R.
Jennison, Phys. Rev. Lett. {\bf 85}, 1910 (2000).

\bibitem{ovesson}
S. Ovesson, Phys. Rev. Lett. {\bf 88}, 116102 (2002).

\bibitem{venablesbrune}
J. A. Venables and H. Brune, Phys. Rev. B {\bf 66}, 195404 (2002).

\bibitem{nandipati}
G. Nandipati and J. G. Amar, Phys. Rev. B {\bf 73}, 045409 (2006).

\bibitem{evanshanney}
M. R. Evans and T. Hanney, J. Phys. A: Math. Gen. {\bf 38}, R195 (2005).

\bibitem{spitzer}
F. Spitzer, Adv. Math. {\bf 5}, 246 (1970).

\bibitem{majumdar}
S. N. Majumdar, S. Krishnamurthy, and M. Barma, Phys. Rev. Lett. {\bf 81}, 3691
(1998).

\bibitem{kanai}
M. Kanai, J. Phys. A: Math. Theor. {\bf 40}, 7127 (2007).

\bibitem{kaupuzs}
J. Kaupuzs, R. Mahnke, and R. J. Harris, Phys. Rev. E {\bf 72}, 056125 (2005).

\bibitem{chamereis}
A. Chame and F. D. A. Aar\~ao Reis, Physica A {\bf 376}, 108 (2007).

\bibitem{payne}
S. H. Payne and H. J. Kreuzer, Phys. Rev. B {\bf 75}, 115403 (2007).

\bibitem{grynberg}
M. Barma, M. D. Grynberg, and R. B. Stinchcombe, Phys. Rev. Lett. {\bf 70}, 1033
(1993).

\bibitem{ammi}
H. S. Ammi, A. Chame, M. Touzani, A. Benyoussef,
O. Pierre-Louis, and C. Misbah,  Phys. Rev. E , {\bf 71} 041603 (2005).

\bibitem{kandel}
D. Kandel, Phys. Rev. Lett. {\bf 78}, 499 (1997).

\bibitem{hodak}
M. Hodak and L. A. Girifalco, Phys. Rev. B {\bf 64}, 035407 (2001). 

\end{references}
\end{document}